\documentstyle[eqsecnum,epsfig,twocolumn,aps]{revtex}

\begin{document}
\draft \preprint{}

\wideabs{

\title{Evidence for an energy scale for quasiparticle dispersion in
$Bi_{2}Sr_{2}CaCu_{2}O_{8}$ }
\author{P. V. Bogdanov$^1$, A. Lanzara$^1{}^,{}^2$,  S. A. Kellar$^1$,
 X. J. Zhou$^1{}^,{}^2$, E. D. Lu$^2$, W. Zheng$^1$, G. Gu$^3$, K. Kishio$^4$, J. -I. Shimoyama$^4$,  Z. Hussain$^2$, and Z. X. Shen$^1$
}
\address{
$^1$Department of Physics, Applied Physics and Stanford
Synchrotron Radiation Laboratory,\\ Stanford University, Stanford,
CA 94305, USA }
\address{
$^2$Advanced Light Source, Lawrence Berkeley National Lab,
Berkeley, CA 94720 }
\address{
$^3$School of Physics, University of New South Wales, P. O. Box 1, Kensington, NSW, Australia 2033}
\address{
$^4$Department of Applied Chemistry, University of Tokyo, Tokyo,
113-8656, Japan}

\maketitle

\begin{abstract}
Quasiparticle dispersion in $Bi_{2}Sr_{2}CaCu_{2}O_{8}$ is
investigated with improved angular resolution as a function of
temperature and doping. Unlike the linear dispersion predicted by
the band calculation, the data show a sharp break in dispersion at
$50\pm10$ $meV$ binding energy where the velocity changes by a
factor of two or more. This change provides an energy scale in the
quasiparticle self-energy.  This break in dispersion is evident at
and away from the d-wave node line, but the magnitude of the
dispersion change decreases with temperature and with increasing
doping.
\end{abstract}

}

\narrowtext

In a conventional metal the observation of an energy scale often
provides significant insight into the physical process in the
material.  The most noted example is the observation of the phonon
anomalies in strong coupling superconductors such as lead, which
had a far-reaching impact on the understanding of the
superconductivity mechanism
\cite{MacMillan,Scalapino2,Scalapino3}. For the high-temperature
superconductors, a peculiar normal state property is the fact that
there appears to be no energy scale, which is often referred to as
the marginal Fermi liquid behavior \cite{Varma}. This behavior is
highly anomalous as one would expect certain energy scales in the
problem, say phonons which are obviously present in the crystal.
In the theoretical context this lack of an energy scale is
believed to be a key feature of a near-by quantum critical point
\cite{qcritical}. In the superconducting state, on the other hand,
there are energy scales observed in the cuprate superconductors.
One of them is the superconducting gap and the other is the
so-called 41 meV magnetic resonance \cite{Rossat,Fong,Arai,Dai}.
The latter has been attributed to the superconducting gap, or the
$\pi$-resonance of the SO(5) theory \cite{Fong,Zhang}.

With its ability to measure both the real and imaginary parts of
the self-energy, $\Sigma(\omega,k)$, angle-resolved photoemission
(ARPES) experiments provide a unique opportunity to further
explore this issue as any relevant energy scale present will
manifest itself in the quasiparticle dynamics.  In the known case
of electron-phonon interaction the coupling causes a kink in the
dispersion and also a change in quasiparticle lifetime near the
phonon energy \cite{Scalapino3}. These canonical changes reveal
effects in the real and imaginary parts of the self-energy due to
the electron-phonon interaction, an effect which is experimentally
observed recently \cite{MO}.  In this letter, we present
high-resolution ARPES data from $Bi_{2}Sr_{2}CaCu_{2}O_{8}$
superconductors as a function of doping and temperature.  We have
observed a clear break in the quasiparticle dispersion near
$50\pm10$ $meV$ binding energy (BE), that results in a change in
the quasiparticle velocity up to a factor of two or more.  This
effect is enhanced in the underdoped sample, and appears to
persists above $T_{C}$ where the break becomes rather broad.
Because the electronic structure calculation \cite{LDA} predicts a
linear dispersion in this range, this result represents an
important effect in the real part of the self-energy with a scale
near $50\pm10$ $meV$. Further, we found that this effect is
present at various points of the momentum space. We believe the
doping, temperature and $\vec{k}$-dependent information presented
here will put a constrain on microscopic theory.

Angle-resolved photoemission data have been recorded at beamline
$10.0.1.1$ of the Advanced Light Source utilizing $22$ $eV$, $33$
$eV$ and $55$ $eV$  photon energies, in a similar set-up as we
have reported recently \cite{Zhou}. The momentum resolution was
$\pm0.1$ degrees,  which is about an order of magnitude better
than our previous study of this material, making the results
reported in this letter possible. The energy resolution was $14$
$meV$. The vacuum during the measurement was better than
$4\cdot10^-{}^1{}^1$ $torr$. The underdoped (UD)
$Bi_{2}Sr_{2}CaCu_{2}O_{8}$  ($Tc = 84K$) and the slightly
overdoped (OD) $Bi_{2}Sr_{2}CaCu_{2}O_{8}$ sample ($Tc = 91K$)
were grown using floating-zone method. The single crystalline
samples were oriented and cleaved in situ at low temperature.

Fig.1a) shows raw ARPES data collected along the $(0,0)$ to
$(\pi,\pi)$ (nodal) direction of the Brillouin zone from the OD
sample at 30K. In panel 1b) we plot the dispersion determined from
the fits to the momentum distribution curves (MDCs) - angle scans
at a constant binding energy \cite{Valla}. MDC plots show a peak
on a constant background that can be fitted very well with a
simple Lorentzian, as illustrated in the inset b2). Error bars in
$k_{\|}$ and energy are determined from the fit uncertainty and
energy resolution respectively.  The data clearly show a feature
dispersing towards the Fermi energy with an obvious break in the
slope near $50$ $meV$ BE.  A similar break in the dispersion was
also observed at photon energies $22$ $eV$ and $55$ $eV$. Data for
all three photon energies is plotted in the inset b1) in panel b).
To describe the dispersion in the range of ($-200$ $meV$ to $0$
$meV$) one needs only two straight lines intersecting near $50$
$meV$ BE. This behavior is clearly different from what one expects
from the LDA or any other electronic structure calculation where a
linear dispersion in this energy range is predicted. Raw MDCs are
plotted in panel 1c), while raw energy distribution curves (EDCs)
are plotted in panel 1 d) for reference.

We present in Fig. 2 (a-c) the dispersions obtained from different
cuts parallel to the $(0,0)$ to $(\pi,\pi)$ direction across the
Fermi surface for the UD sample at 20K. Within the error bars, the
data are again well described by two straight lines with a break
near $50$ $meV$ BE. The energy position of the break is constant
throughout the BZ within the experimental uncertainty, despite the
opening of the gap. Fig. 3 shows the locations in the
two-dimensional zone where the break is experimentally observed.
It demonstrates that the effect is present in all directions.

We have investigated this effect as a function of doping and
temperature. The effect appears to be stronger in the underdoped
sample. The change of the quasiparticle velocity at the break is
different, which can be illustrated by data along the $(0,0)$ to
$(\pi,\pi)$ direction. For the underdoped sample, the
quasiparticle velocity determined from the MDC fits shows a break
from $3.6$ $eV\AA$ at higher binding energies to $1.5$ $eV\AA$
near the Fermi level. The respective velocities for the optimally
doped samples are $2.6$ $eV\AA$ and $1.6$ $eV\AA$. The error in
velocities from the fits is $\pm.1 eV\AA$.  Main source of error
is the possible misalignment \cite{comment}, thus causing some
uncertainty in quantitative results. However, the general trend
discussed above is robust.

In general one expects to see complementary effects in dispersion
and EDC and MDC peak widths as they reflect the quasiparticle
self-energy. The self-energy can be easily extracted from an ARPES
experiment if $Im\Sigma(\vec{k},E)$ is much smaller than the
energy. In this case MDC and EDC methods give the same result for
the peak position and for the peak width interpretable as $Re$ and
$Im$ parts of the self energy respectively. In high Tc's
extracting the self energy from ARPES is harder because the EDC
peak energy is comparable to the peak width for $E \geq 30 meV
BE$. However, assuming weak k-dependence of the
$Im\Sigma(\vec{k},E)$ \cite{Varma}, the deviation of the MDC
dispersion from the LDA calculation gives real part of the self
energy and the MDC peak widths represent imaginary part
\cite{Bogdanov}. Fig.2 (a1-c1) shows MDC widths in momentum space
along various cuts. The corresponding energy width is given by the
momentum width of the MDC peak multiplied by the velocity if the
scan direction is along the energy gradient direction. In our
geometry this condition is satisfied only along the nodal
direction. In Figure 2 d) we plot the energy width from MDC
together with EDC width. The step effect in MDC energy width is
due to linear approximation to the dispersion in determining the
velocity, smoother transition is expected for a less dramatic
behavior. EDC peak widths do not simply give $Im\Sigma(\vec{k},E)$
in the case of broad peaks. Furthermore, EDC data is complicated
by energy-dependent background and Fermi cut-off. Nevertheless,
the EDC data still indicate a more abrupt change in the width at
the energies corresponding to the kink in the dispersion, as shown
in Fig. 2 (d-f). We feel that the qualitative consistency between
MDC and EDC results is sufficient to make the case for the strong
self energy effect in the data. The lack of quantitative agreement
between EDC and MDC is a manifestation of the subtle lineshape
issues discussed above.

The dispersions determined from the OD sample above $T_{C}$ along
the $(0,0)$ to $(\pi,\pi)$ ($\Gamma-Y$) direction are shown in
Fig. 4a), while the low temperature dispersions ($T<T_{c}$) are
reported in 4b) . The dispersions exhibit the same break structure
as contrasted to the straight line. The change of dispersion is
more difficult to see in the high temperature data compared to low
temperature data, but a weak residual effect still appears to be
present. In Fig. 4c)we show the temperature dependence of the EDC
width. We see a clear change in $2\Gamma$ around $50\pm10 meV$ in
the low temperature data, but the effect is harder to see above
Tc.

We now discuss the origin of the strong self-energy effect near
$50\pm10$ $meV$.  The first possibility that comes to mind is the
electron-phonon interaction as there are phonons of this energy
scale in the compound \cite{phonons}. This would explain the
persistence of the feature throughout the Brillouin zone and the
persistence of the feature above the superconducting transition
temperature, since phonons are $\vec{k}$ and T independent.
However, as shown recently \cite{MO}, the dispersion tends to
recover to the one-electron result when the energy is well above
the typical phonon energies, with the total range of the perturbed
dispersion below $E_{F}$ equal to half the Debye temperature. In
our case, there is no indication that the dispersion will recover
to LDA behavior and the high energy part of the data cannot be fit
with a line passing through the Fermi surface crossing. Another
problem with the phonon scenario is that it is not a natural
explanation on why the effect is stronger in underdoped sample.

The second and in our view most likely possibility is the electron
coupling to collective magnetic excitations
\cite{Rossat,Fong,Arai,Dai,Zhang}. The neutron mode is found below
$T_{C}$ near $(\pi,\pi)$ at $41$ $meV$ with a width of $0.6\pi$ in
$Bi_{2}Sr_{2}CaCu_{2}O_{8}$ system \cite{FongNature}. The energy
scale of the neutron mode is consistent with the $50\pm10$ $meV$
feature seen in our experiments.  This picture is also consistent
with the fact that the underdoped sample shows stronger effect
than the overdoped one \cite{FongNature,He}.  The broadness of the
neutron peak in this compound makes it possible to explain the
persistence of the effect throughout the Brillouin zone. Mode's
energy decrease in overdoped sample and the fact that the mode is
mainly seen below $T_{C}$ also supports this interpretation of our
data. However, we caution that the exact temperature where this
mode turns on is a sensitive function of doping and impurities
\cite{Mook}. Of course, we cannot rule out the possibility of a
combination of effects due to phonons and magnetic excitations
because both have similar energy scales. We should note that the
coupling of a quasiparticle to collective excitations was
previously discussed, but in a very different context
\cite{Shen,Norman}. The break in quasiparticle dispersion of
optimally doped sample along the nodal direction is also present
in the data of Valla $et al.$\cite{Valla}. However, these authors
did not elaborate on this issue and they suggested the absence of
energy scale in the problem.

The third possibility is that we see the effect related to the
opening of the superconducting gap which is of the order of
$50\pm10$ $meV$ in this compound.  Because the anti-nodal
direction near $(\pi,0)$ has a very high density of states, we
expect an effect when this energy scale is reached \cite{Bishop}.
The down side of this scenario is that we do not have theoretical
calculations on specifics of the effect.  At this point, we do not
feel that we can rule out possible interpretation of the data
based on the stripe scenario
\cite{Emery,Zaanen,Tranquada,Bianconi,Zhou} where the 50meV kink
in the dispersion reflects the characteristic frequency of the
fluctuating stripes.

In summary, we have studied doping, momentum and temperature
dependence of the quasiparticle dispersion using ARPES with very
high momentum resolution.  We have uncovered an energy scale of
$50\pm10$ $meV$ where the quasiparticle dispersion shows a strong
change.  This effect is seen in all directions, and is stronger in
the underdoped sample.  The effect is strongest in data below
$T_{C}$.  We expect that the data presented here is an important
part of the puzzle related to high$-T_{C}$ superconductivity.

We would like to thank J. D. Denlinger for the help with data
analisys software. We would like to thank P. D. Johnson,B. O.
Wells,  A. Fedorov, T. Valla, S.C. Zhang, D.J. Scalapino, Steve
Kivelson, D.H. Lee, Bob Laughlin, and P.A. Lee for discussion. The
experiment was performed at the Advanced Light Source of Lawrence
Berkekely National Laboratory. The Stanford work was supported by
NSF grant through the Stanford MRSEC grant and NSF grant
DMR-9705210.  The work at ALS was supported by DOE' Office of
Basic Energy Science, Division of Materials Science with contract
DE-AC03-76SF00098.  The SSRL's work was also supported by the
Office's Division of Materials Science.

\begin{figure}
 \leavevmode
 \epsfxsize=\columnwidth
 \epsfig{file=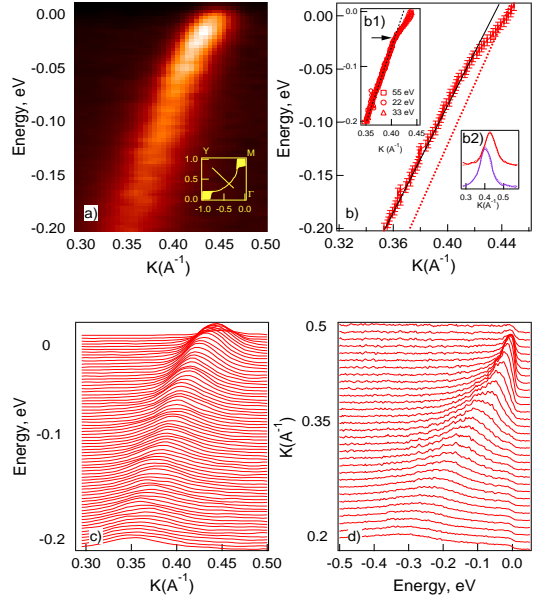,clip=,angle=0,width=3in,height=3.3in}
 \vspace{0in}
\caption{Panel a) shows raw data obtained using Scienta angle mode
for slightly overdoped (Tc=91K) $Bi_{2}Sr_{2}CaCu_{2}O_{8}$ along
nodal direction ($\Gamma-Y$) of the BZ at 33 eV photon energy. The
position of the cut is given in the inset. Panel b) shows the
dispersion of the quasiparticle determined from the MDC fits of
the data in panel a). The theoretical dispersion from LDA
calculation is also included (dotted straight line). Energy is
given relative to the Fermi energy. Inset b1) shows the dispersion
along this direction obtained at $22 eV$, $33 eV$ and $55 eV$.
Inset b2) shows MDC's at $16$ (blue) and $55$ (red) $meV$ BE.
Dashed lines represent Lorenzian fits. Panels c) and d) show raw
MDCs and EDCs respectively. }
\end{figure}

\begin{figure}
 \leavevmode
 \epsfxsize=\columnwidth
 \epsfig{file=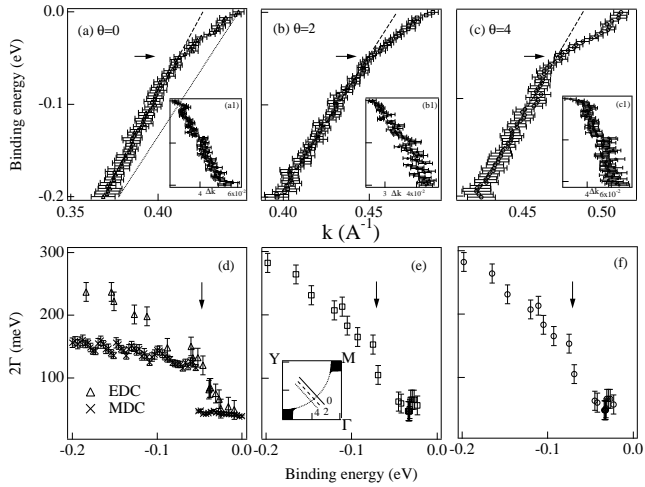,clip=,angle=0,width=3.5in,height=2.6in}
 \vspace{0in}
\caption{Panels a) to c) show the MDC-derived dispersions for the
underdoped $Bi_{2}Sr_{2}CaCu_{2}O_{8}$ ($T_{C}=84K$) for cuts
parallel to $\Gamma-Y$ direction vs the momentum. (The positions
of the cuts in the BZ are shown in the inset of Fig. 2e).) The
linear fits to the dispersions are also shown. Insets a1) to c1)
show MDC derived quasiparticle widths in momentum space along the
cuts as a function of binding energy. Panels e) and f) show EDC
derived quasiparticle widths in energy space along the cuts as a
function of binding energy. Panel d) shows EDC width together with
the peak width in energy space derived from MDC of inset a1) via
velocity determined from dispersion of panel a).Energy is given
relative to Fermi energy.}
\end{figure}

\begin{figure}
 \leavevmode
 \epsfxsize=\columnwidth
 \epsfig{file=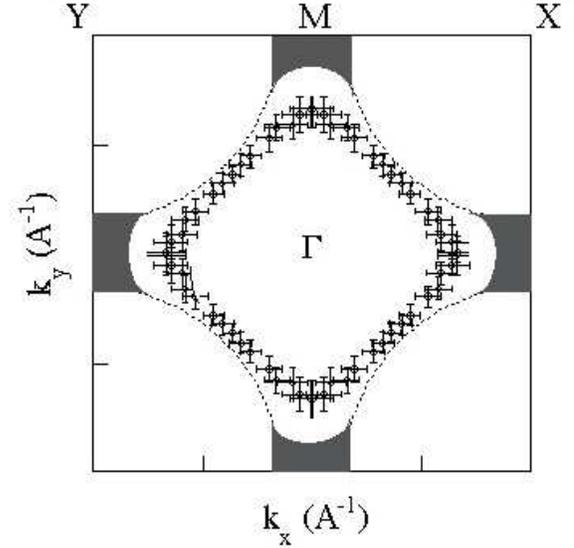,clip=,angle=0,width=3in,height=3in}
 \vspace{0in}
\caption{Kink position as a function of $\vec{k}_x$ and
$\vec{k}_y$ is plotted in the BZ (circles). Eight-fold
symmetrization procedure was applied. Error bars reflect
uncertainty in kink position from the MDC fits and the
experimental angular resolution perpendicular to the scan
direction. Fermi surface is plotted for reference (dashed lines).}
\end{figure}

\begin{figure}
 \leavevmode
 \epsfxsize=\columnwidth
 \epsfig{file=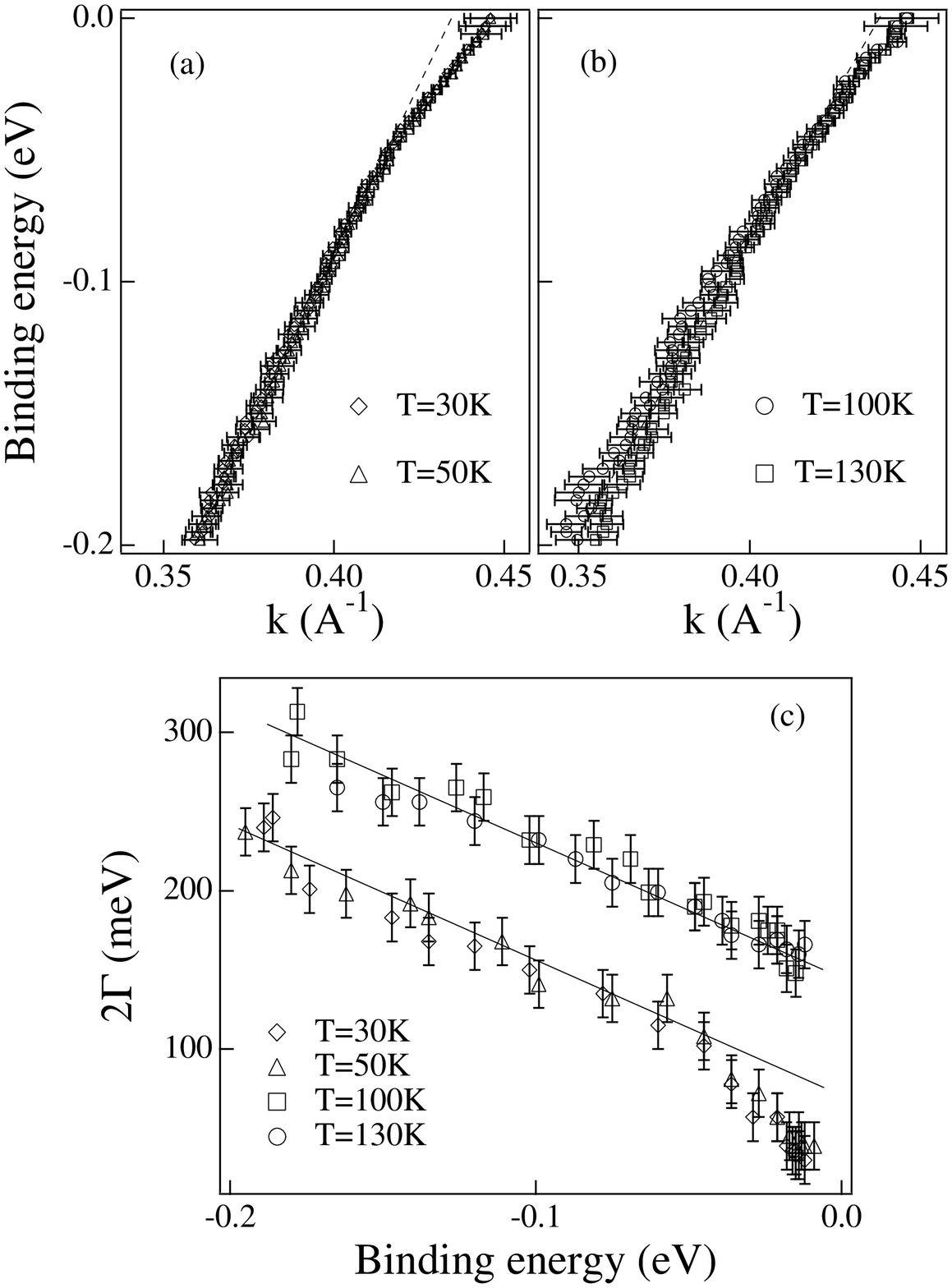,clip=,angle=0,width=3in,height=3.7in}
 \vspace{0in}
\caption{In Panel a) dispersion in the slightly overdoped
$Bi_{2}Sr_{2}CaCu_{2}O_{8}$ ($T_{C}$ = $91K$) along nodal
direction ($\Gamma-Y$) of the BZ below $T_{C}$ is reported. Dotted
line represents linear fit into the high energy part of the
dispersion. Energy is given relative to Fermi energy. In Panel b)
dispersion above $T_{C}$ is reported for the same sample. Dotted
line represents linear fit into the high energy part of the
dispersion. Energy is given relative to Fermi energy. Panel c)
shows EDC derived quasiparticle widths of the spectral feature as
function of binding energy. High temperature data is shifted up by
100 $meV$ for clarity. Energy is given relative to Fermi energy.}
\end{figure}


\begin{references}
\bibitem {comment} In figure 2 we plot dispersions as a
function of angle. The change of velocities with angle is up to $1
eV\AA$ per degree. Although our angular resolution is $\pm.1$
degrees, the error in absolute alignment is $\pm1$ degree, which
is the main source of uncertainty. However, for both dopings we
sampled over the whole Brillouin zone. This gives us internal
reference point for alignment, since velocities decrease as we go
away from nodal direction. We have measured two overdoped samples
at three photon energies. The velocities we determined for both
samples for low binding energy part are within $.1 eV\AA$. There's
no detectable change in velocity with doping or photon energy in
this region. Other groups reported $1.14 eV\AA$\cite{Valla} and
$1.6 eV\AA$ \cite{ArgonneVelocity} for the low binding energy
velocities. For the high energy part the difference with samples
is also negligible. For the sample we measured at three photon
energies we find $2.6\pm.1 eV\AA$. For a sample of the same doping
other experiment gave $2.5 eV\AA$.
\end{references}
\end{document}